# Giant and bidirectionally tunable thermopower in non-aqueous ionogels enabled by selective ion doping

Sijing Liu[1,†], Yuewang Yang[1,†], He Huang[1], Jiongzhi Zheng[1], Gongze Liu[1], Tsz Ho To[1], Baoling Huang[1,2,*]

[1]Department of Mechanical and Aerospace Engineering, The Hong Kong University of Science and Technology, Clear Water Bay, Kowloon, Hong Kong SAR, China.

[2]The Hong Kong University of Science and Technology Foshan Research Institute for Smart Manufacturing, Clear Water Bay, Kowloon, Hong Kong SAR, China.

*Corresponding author. Email: mebhuang@ust.hk

†These authors contributed equally to this work.

## Abstract

Ionic thermoelectrics show great potential in low-grade heat harvesting and thermal sensing owing to their ultrahigh thermopower, low cost and ease in production. However, the lack of effective n-type ionic thermoelectric materials seriously hinders their applications. Here, we report giant and bidirectionally tunable thermopowers within an ultrawide range from -23 to +32 mV K-1 at 90% RH in solid ionic-liquid-based ionogels, rendering it among the best n- and p-type ionic thermoelectric materials. A novel thermopower regulation strategy through ion doping to selectively induce ion aggregates via strong ion-ion interactions is proposed. These charged aggregates are found decisive in modulating the sign and enlarging the magnitude of the thermopower in the ionogels. A prototype wearable device integrated with 12 p-n pairs is demonstrated with a total thermopower of 0.358 V K-1 in general indoor conditions, showing promise for ultrasensitive body heat detection.

**Teaser**

The giant thermopower of an ionic-liquid-based ionogel can be tuned within an ultrawide range by selective ion doping.

**Introduction**

More than 60% of total energy consumption in the world is eventually lost as waste heat (*1*). Most of waste heat is identified as low grade heat below 100 °C, making recycling difficult (*2*). Thermoelectrics (*3*) based on the Seebeck effect, in which a thermovoltage $V$ is produced under a temperature difference $\Delta T$, provide a feasible and simple solution to converting heat to electricity directly, empowering them with the ability of thermal sensing (*4*) or heat harvesting. Common high-performance thermoelectric materials with either electrons or holes carrying both charges and heat, have Seebeck coefficients (or thermopower, $\alpha = -V/\Delta T$) ranging from only tens of to a few hundred microvolts per Kelvin (*5*). The limited Seebeck coefficients restrain their thermoelectric performance in both power generation and thermal sensing applications; however, it is challenging to significantly improve the Seebeck coefficient without sacrificing other important thermoelectric properties such as electrical conductivity due to their strong intercoupling.

In the past few years, giant thermopowers of the order of 10 mV K$^{-1}$ have been reported frequently in ionic conductors (*6–12*). Very similar to their electronic siblings, a thermovoltage will be built up when ions in an ionic conductor thermodiffuse from the hot side to the cold side under a temperature gradient (*13*) (also called Soret effect (*14*)). However, the ions can only accumulate at the electrodes instead of entering the external circuit, so there is no constant current generated (*15*). The ionic thermopower $\alpha_i$ is also derived from the ratio of the open circuit voltage over the temperature difference, i.e., $\alpha_i = -V/\Delta T$. Generally, an ionic thermoelectric material is p-type with a positive thermopower if the cold electrode has a higher electrical potential, which indicates the cations thermodiffuse faster than anions, and vice versa.

The ionic thermopower is correlated to the ion transport entropy in ionic thermodiffusion (*16*), which is determined by interactions between ions and the local environment. The ionic thermopower in aqueous and organic solutions have been intensively investigated in the last few years and thermopowers up to ~10 mV have been reported (*6*, *7*, *16*). Recently solid electrolytes of giant ionic thermopowers (*8–10*), especially polymer gels, attract great attention due to their advantages in manufacturing and device integration. The introduction of solid matrix can also significantly enhance the ionic thermopower through the strong ion-matrix interactions. Polymer ionogels based on non-volatile ionic liquids (ILs) are less explored compared with hydrogels and preferred in many potential applications due to their great stability. A record high thermopower (+26 mV $K^{-1}$) has been achieved in poly(vinylidene fluoride-co-hexafluoropropylene) (PVDF-HFP) based ionogels, which is attributed to the ion-dipole interaction between PVDF-HFP and the IL (*11*). Interestingly, it is recently reported that by changing the polymer composition, the ionic thermopower can be tuned from -4 mV $K^{-1}$ to +14 mV $K^{-1}$ in an IL-based polymer electrolyte (*12*). So far, all the strategies to tune the sign and magnitude of ionic thermopowers have been focused on modulating the interactions between ions and the matrix environment. However, owing to various considerations such as the stability and mechanical strength, suitable polymer matrices for electrolytes are limited and the regulation of the interactions between the selected polymer matrix and ions is also challenging. Despite the significant advances in achieving giant ionic thermopowers, most of state-of-the-art ionic thermoelectric materials are p-type or have positive ionic thermopowers. Up until now, n-type ionic thermoelectric materials are scarce and the relatively low thermopower can hardly match the performance of p-type ionic thermoelectric materials. The lack of high-performance n-type materials seriously hampers the adoption of ionic thermopiles in practical applications.

Here, we propose a novel strategy in which strong ion-ion interactions are selectively introduced into the ionic thermoelectric materials to tune the ionic thermopower in desired positive or negative directions. We adopt a binary IL-based polymer ionogel as the start, which is flexible and can be manufactured using low-cost solution processes (*17*). When ionic salts that can closely interact with original ions are added to form ternary polymer ionogels, the ionic thermopower can be drastically tuned in either positive or negative directions within the range from -15 mV K$^{-1}$ to +17 mV K-1 at 60% RH. We achieve a giant negative thermopower of -15 mV K-1 using this strategy, reaching a record value for n-type ionic thermoelectric materials under the same condition. The thermopower is even magnified to -23 to 32 mV K-1 at high RH (90%). Through various characterization techniques including Raman, infrared and nuclear magnetic resonance (NMR) spectroscopies, we conclude that the bidirectionally tunable thermopower results from interactions between ions of the salts and those of the IL. We also demonstrate a flexible wearable ionic thermoelectric device consisting of 12 p-n pairs based on the developed p- and n-type ionogels, achieving a device thermopower of 358 mV K$^{-1}$.

**Results**

**Tunable ionic thermopower in ternary polymer ionogels**

We chose a common binary IL-based polymer ionogel as a start for demonstration, which adopts the copolymer poly(vinylidene fluoride-co-hexafluoropropylene) (PVDF-HFP) as a solid polymer matrix and the IL 1-ethyl-3-methylimidazolium ([EMIM]) bis(trifluoro-methylsulfonyl)imide ([TFSI]) electrolyte. The pure EMIMTFSI/PVDF-HFP ionogel with a weight ratio of $W_{IL}:W_{polymer}$=4:1 shows an ionic thermopower of around -4 mV K$^{-1}$ (60%

RH, room temperature), agreeing quite well with the result reported in literature (*12*). The negative sign of the thermopower in the pure EMIMTFSI/PVDF-HFP ionogel indicates anions move faster than cations under a temperature gradient, owing to the interactions of ions and the high polarity fluorine-containing polymer chains (*12*). Since the ion motions are strongly influenced by both their ambient polymer matrix and surrounding ionic environment, it is expected that adding specific salts into the binary ionogel can regulate the ionic environment and in turn selectively tune the ion interactions and diffusivity difference in the system, consequently altering the sign and magnitude of the ionic thermopower.

The preparation of ternary polymer ionogels with PVDF-HFP polymer matrix is similar to the process for the binary EMIMTFSI/PVDF-HFP ionogel, except that specific ionic salts are added to EMIMTFSI to form an IL solution before the mixing of the IL with PVDF-HFP solution. In the ternary polymer gels consisting of PVDF-HFP, ILs and salts, the polymer is a host of the liquid electrolyte, and the IL acts as a liquid solvent for salts as well as an ion provider. Lithium tetrafluoroborate ($LiBF_4$), which is a widely used lithium salt for electrolyte in lithium-ion batteries (*18*), is chosen for ion doping in the polymer gel to increase the magnitude of the negative ionic thermopower, as $Li^+$ cations preferentially coordinate with the $TFSI^-$ anions when dissolved in ILs (*19*). After doping 0.5 M $LiBF_4$ in the EMIMTFSI/PVDF-HFP ionogel, the ionic thermopower of the polymer ionogel is boosted from -4 to -15 mV $K^{-1}$ (Fig. 1A and Fig. S3), which is much higher than the previous value (-4 mV $K^{-1}$) for state-of-the-art n-type ionic thermoelectric materials and matches with the best performance of p-type ones (Fig. 1B). On the other hand, as halide ions tend to interact with $EMIM^+$ cations of the ILs, the thermodiffusion of the electrolyte ions shifts to a cation-dominant direction after halide salt doping. An n- to p-type conversion in the EMIMTFSI/PVDF-HFP ionogel with the ionic thermopower changing from -4 to +17 mV $K^{-1}$ is attained by adding 0.5 M 1-ethyl-3-methylimidazolium chloride (EMIMCl) (Fig 1A and Fig. S3). The positive thermopower of +17 mV K$^{-1}$ achieved with ionic salt doping is slightly higher than that (+14 mV K$^{-1}$) obtained in EMIMTFSI/PVDF-HFP ionogel with non-ionic PEG additive. The role of the added salts will be discussed in detail in later sections.

Figure 1C shows the thermopower of the ternary ionogels doped with 0.5 M salts in IL electrolyte but with different electrolyte to polymer weight ratios in general conditions. When there is no PVDF-HFP matrix, the 0.5 M EMIMCl/EMIMTFSI and 0.5 M LiBF$_4$/EMIMTFSI liquid electrolytes show thermopowers of +5.8 and -5.5 mV K$^{-1}$, respectively. Incorporating the liquid electrolytes into the PVDF-HFP matrix retains the signs of the thermopower but greatly boosts their magnitudes, reaching around +17 mV K$^{-1}$ for 0.5 M EMIMCl/EMIMTFSI/PVDF-HFP and -15 mV K$^{-1}$ for 0.5 M LiBF$_4$/EMIMTFSI/PVDF-HFP ionogels with only ~10 wt% PVDF-HFP content. This implies that the PVDF-HFP in the ionogels acts more than as an inert matrix to constrain the liquid portion. Apparently, the ion-polymer chain interaction can greatly boost the thermopower, as also observed in the literature (*9–12*). It is interesting to note that when the liquid electrolytes form ionogels with PVDF-HFP, the thermopower is almost independent on the electrolyte/polymer weight ratio within the range considered. This indicates that small polymer fraction is sufficient to apply the influence on the ion thermodiffusion. It is conjectured that the PVDF-HFP matrix does not change the dominating ion species but lowers the mobility of all ion species (*20*) and results in a larger mobility difference between the cations and anions, which contributes to larger absolute values of the thermopowers.

Considering the hydrophobic ionic liquid is still hygroscopic (21) and meanwhile, the addition of lithium salts or chloride salts further increases hygroscopicity of the ionogel, thermopowers were also measured in the relative humidity range from 40% to 90% (Fig.

1D). At 90% RH, the thermopower of the 0.5 M LiBF4/EMIMTFSI/PVDF-HFP and 0.5 M EMIMCl/EMIMTFSI/PVDF-HFP ionogels reaches -23 and +32 mV K-1 respectively, ranking the highest thermopowers reported so far. In a moderate relative humidity environment (60% RH), which is adopted all through this manuscript unless specified, the thermopower decreases to -15 and +17 mV K-1. A continuing decrease to -11 and +6.7 mV K-1 is observed in the dry environment (40% RH). The thermopower of the 0.5 M EMIMCl/EMIMTFSI/PVDF-HFP ionogel shows larger dependence on the relative humidity than the 0.5 M LiBF4/EMIMTFSI/PVDF-HFP ionogel. And the water effects like an amplifier on the thermopower. To be more specific, the magnitude increases with higher relative humidity while the sign maintains unaltered.

Tuning negative thermopower by LiBF4 doping

As discussed above, LiBF4 shows good performance in enhancing the negative thermopower of the EMIMTFSI/PVDF-HFP ionogel. For the following measurements, the weight ratio of liquid electrolyte to polymer was controlled at 4:1. The salts were pre-dissolved in the IL to attain various concentrations. Figure 2A and Figure S4A exhibit the thermopowers and ionic conductivities of LiBF4/EMIMTFSI/PVDF-HFP ionogels with different LiBF4 concentrations at 60% RH. The ionic conductivity of the ionogel gradually increases from 2 mS cm-1 to around 4 mS cm-1 when the concentration of $LiBF_4$ increases from 0 to 1 M. Upon the addition of $LiBF_4$, the thermopower of the ionogel first increases from -4 mV K$^{-1}$ to a maximum value -15 mV K$^{-1}$ when the concentration of $LiBF_4$ increases to around 0.5 M and then slightly falls back to -11 mV K$^{-1}$ at 1 M $LiBF_4$. This drop might be due to the increase of viscosity of the liquid electrolyte at higher lithium salt concentration (*22*).

At the microscopic level, the addition of lithium salts hinders the thermodiffusion of IL cations and favours that of the anions by establishing ion interactions with the IL. $Li^+$ cations tend to form ion clusters with the anions in an IL/lithium salt mixed electrolyte, as evidenced by experiments (*19*) and simulations (*23*), which diffuse in the form of $[Li(TFSI)_n]^{1-n}$ clusters if the anion is $TFSI^-$. The formation of lithium clusters in the $LiBF_4$/EMIMTFSI/PVDF-HFP ionogels is confirmed with Raman spectroscopy (Fig. 2B). The band at 742 $cm^{-1}$ is assigned to free $TFSI^-$ anions for the pure IL ionogel, and as the concentration of $LiBF_4$ increases, a new band grows at 748 $cm^{-1}$, indicating that the $TFSI^-$ anions form aggregates with $Li^+$ cations (*19*). Similarly, the band at 765 $cm^{-1}$ is assigned to free $BF_4^-$ anions (*24*) and this band becomes wider with higher $LiBF_4$ concentration due to more paired $BF_4^-$ anions locating at 770 $cm^{-1}$. These results confirm $[Li(BF_4)_m(TFSI)_n]^{1-m-n}$ clusters formed in the mixed-anion $LiBF_4$/EMIMTFSI/PVDF-HFP ionogel. The lithium clusters have relatively larger volumetric size compared to single ions, leading to more sluggish thermodiffusion. In this case, the lithium clusters effect as many almost immobilized negative charge centers and exert a drag effect on $EMIM^+$ cations due to Colombian force, thus slowing down their movements; meanwhile, the remaining $TFSI^-$ anions can move freely in the ionogel, resulting in a larger negative thermopower (*10*). A schematic diagram to demonstrate this mechanism is shown in Fig. 2C.

We find that the formation of lithium clusters is essential for the enhanced negative thermopower. As a control experiment, the 0.5 M $LiBF_4$ in the ionogel was substituted with 1-ethyl-3-methylimidazolium tetrafluoroborate ($EMIMBF_4$) of the same molar concentration so that the $Li^+$ cations could be excluded while $BF_4^-$ anions were retained. The measured ionic thermopower of the resulting ionogel is -3.7 mV $K^{-1}$ (Fig. 2D), almost equal to that of the pure EMIMTFSI/PVDF-HFP ionogel. Besides, the boost of the negative thermopower is a synergic effect of both $Li^+$ cations and $BF_4^-$ anions. To investigate the effect of $BF_4^-$ anions, lithium bis(trifluoromethanesulfonyl)imide (LiTFSI), which has the same anion as the IL, was selected as the dopant in the ionogel. The 0.5 M LiTFSI/EMIMTFSI/PVDF-HFP ionogel shows a thermopower of -7.6 mV $K^{-1}$. This value is slightly higher than that of pure EMIMTFSI/PVDF-HFP ionogel but much lower than that of the 0.5 M $LiBF_4$/EMIMTFSI/PVDF-HFP ionogel. $BF_4^-$ anions exhibit larger polarity than $TFSI^-$ anions (*25*), leading to stronger coordination with $Li^+$ cations. The larger $[Li(BF_4)_m(TFSI)_n]^{1-m-n}$ clusters with more negative charges are suspected to contribute to higher ionic thermopowers than the monoanion case. These two comparisons coherently confirm the strong ion-ion interactions introduced by $LiBF_4$ in the IL ionogel.

**Reversed positive thermopower by adding EMIMCl**

Halide ions can change the sign of the thermopower of the IL ionogel. The thermopower and ionic conductivity of EMIMCl/EMIMTFSI/PVDF-HFP ionogels containing various concentrations of EMIMCl at 60% RH were also investigated. It can be seen in Fig. 3A that with the addition of EMIMCl, the thermopower first changes from -4 mV $K^{-1}$ to 0 and then further increases until it reaches a peak value as high as +17 mV $K^{-1}$. The ionic conductivity increases with EMIMCl addition at the beginning and then remains almost unchanged as the concentration of EMIMCl further increases.

The interactions between EMIMCl and the IL were studied by $^1$H NMR and Fourier-transform infrared (FTIR) spectroscopy in the liquid state and ionogels, respectively. For $^1$H NMR measurements, different concentrations of EMIMCl/EMIMTFSI were dissolved in d6-acetone and measured. The entire $^1$H NMR spectra of the samples together with the peak assignment can be found in Fig. S5, where the pure EMIMTFSI spectra are consistent with the literature results (*26*). Upon the increase in the concentration of EMIMCl, all peaks remain at their positions except for the peak assigned to the hydrogen in C(2)-H, the most acidic proton in the imidazolium ring (*27*), which gradually shifts from 9.05 ppm to 9.49 ppm, as shown in Fig. 3B. This peak shift indicates the formation of hydrogen bonds between C(2)-H in EMIM$^+$ cations and Cl$^-$ anions, which is further confirmed by FTIR spectra (Fig. 3C). FTIR spectra of the IL based ionogels from 3050 to 3200 cm$^{-1}$ are shown, because this region contains characteristic features of the imidazolium ring. For the neat IL ionogel, the peak at 3126 cm$^{-1}$ is assigned to the C(2)-H stretch (*28*). The red shift after adding EMIMCl is a good indicator of hydrogen bonds between the cation and Cl$^-$ anion (*27*). The $^1$H NMR and FTIR results can support and verify the hydrogen bonds formation between the EMIM$^+$ cations and Cl$^-$ anions.

Usually in imidazolium halide ILs, the cations and anions connect each other to form a hydrogen-bonded network (*29*). In the ionogel consisting of EMIMTFSI and EMIMCl mixture where there are abundant EMIM$^+$ cations and limited Cl$^-$ anions, the network will break into Cl$^-$ anion-centered aggregates as illustrated in Fig. 3D. In the thermodiffusion, similar to the previously mentioned lithium clusters, these ion aggregates with positive charges slow down surrounding anions, leaving the movement of cations dominating. Thus, the addition of Cl$^-$ anions can reverse the thermopower to a positive value. To verify the effect of ion aggregates from hydrogen bonding, we substitute the chloride with 1-ethyl-3-methylimidazolium bromide (EMIMBr), in which Br$^-$ anions have comparative abilities on hydrogen bonds formation and find similar thermopower (Fig. S6). In summary, the hydrogen bonding based ion aggregates influence the thermodiffusion of original ion species and further tune the thermopower towards a positive direction.

**Flexible thermoelectric device**

Owing to the large positive and negative ionic thermopowers in the ternary ionogels, we combined 12 pairs of the p-type 0.5 M EMIMCl/EMIMTFSI/PVDF-HFP and n-type 0.5 M

LiBF$_4$/EMIMTFSI/PVDF-HFP ionogels to construct a flexible ionic thermoelectric device with a significant total thermopower. As the polymer materials are soft and easy to bend, we can attach the device to the curved surface of human body to detect temperature change and heat flux or harvest body heat energy. Fig. 4A shows the construction of the device by sandwiching a pre-punched 1 mm-thick 3M VHB tape between copper electrodes patterned on polyethylene films. The p- and n-type ionogel solutions were alternately injected into the cavities in the 3M VHB tape and the dried p- and n-type elements were connected in series by the electrodes. A temperature difference was generated along the cross-plane direction using two Peltier chips, and the corresponding thermovoltage between the two electrodes was recorded. The real-time temperature difference and voltage curves are shown in Fig. 4B. The open circuit voltage changes almost synchronously with the temperature difference, showing the fast response of the device with a relaxation time of about 50 s. Through a linear fitting of the thermovoltage with respect to the temperature difference, a total device thermopower of 0.3579 V K$^{-1}$ is obtained at 60% RH (Fig. 4C). The average thermopower of one p-n pair is 29.8 mV K$^{-1}$, very close to the sum of the thermopowers of a p-type (+17 mV K$^{-1}$) and n-type (-15 mV K$^{-1}$) leg pair.

Figure 4D shows that the flexible thin-film device can be adhered to human arms and produce a stable thermovoltage in an indoor environment (25 °C), which can be used to monitor the skin surface heat flux. When the device was worn on the arm, we observed a sharp voltage increase to 0.33 V in 10 s since one side was heated immediately by the warm skin. The voltage then decayed as the temperature propagated from the skin to the top surface of the device until reaching a plateau at around 0.07 V. Even a small wind disturbance at 200 s was recorded by the voltage fluctuation. According to the measured total thermopower, the stable temperature difference across the device was around 0.2 K. The convective skin surface heat flux was then calculated to be 34.8 W m$^{-2}$ based on the thermal conductivity of the ionogels (Table S1). The temperature difference between the air and the skin was measured to be around 8 K and the corresponding convective heat transfer coefficient can therefore be calculated based on the measured heat flux to be 4.35 W m$^{-2}$K$^{-1}$, lying in the proper range for the heat transfer of human body in natural convection (*30*). A higher voltage signal can be expected if we conduct the measurement in a colder environment to enlarge the temperature difference. This wearable device can be used for high-sensitivity body heat flux monitoring in health care applications. For example, a typical 1-mm-thick commercial Hukseflux FHF02SC copper-constantan thermopile heat flux sensor offers a sensitivity of ~5×10$^{-6}$ V/(W/m$^2$) (*30*). In contrast, the sensitivity of this flexible wearable device of the same thickness can reach 1.8×10$^{-3}$ V/(W/m$^2$). Meanwhile, due to the high thermal resistivities of the polymer materials, a large temperature gradient can be retained even if the sensor thickness is small.

Though the thermally induced nonuniform ion distribution cannot directly do work to the external circuit, the generated thermovoltage can be used for heat harvesting by forming a thermoelectric supercapacitor (*7*) as illustrated in Fig. S7. The supercapacitor adopts an electrode-ionogel-electrode layered structure and as the cross-plane temperature difference arouses thermovoltage in the electrolyte, connecting the two electrodes will allow electrons flow through the external circuit to balance the thermovoltage. After heat and the external circuit is removed, the electrodes have been charged by trapped electrons and can discharge once connected again. The electrode properties play a major part in the discharge capacity and CNT electrodes with large specific surface areas can be used to deliver higher energy compared with metal electrodes (Fig. S8A). The ionic thermoelectric capacitors fabricated in this work can generate energy of 1006 μJ m$^{-2}$ with Ag electrodes or 7982.5 μJ m$^{-2}$ with CNT electrodes in half a thermal cycle at a temperature difference of 1 K (Fig. S8B).

## Discussion

Different from previous efforts that focus on modulating the ionic thermopower in ionic conductors through engineering the ion-matrix interactions, here we propose a new tuning mechanism which takes advantage of strong ion-ion interactions in ternary polymer ionogels. We have demonstrated the effective modulation of the ionic thermopower of EMIMTFSI/PVDF-HFP polymer ionogel in both positive and negative directions simply with selective ion doping. With the addition of ion salts into the IL-based polymer ionogel to form ternary polymer electrolytes, the ionic thermopower can be continuously tuned from -23 mV K-1 to +32 mV K-1 at 90% RH, which is the widest tuning range reported so far. We show that through the appropriate selection of doping salts which interact preferentially with the cations or the anions to form ion aggregates, the thermodiffusion of the ions in ionogels can be effectively controlled in a desired manner to tailor the ionic thermopower, while moisture will further amplify the magnitude. It is expected that this ion regulation strategy can be applied in other ionic thermoelectric materials. A prototype flexible ionic thermoelectric wearable device consisting of 12 pairs of the p- and n-type polymer ionogel elements shows a thermopower of 0.3579 V K$^{-1}$. It can sensitively measure the skin heat flux by producing a thermovoltage utilizing the heat of the human body. These results provide a promising solution to reach the demands of future human interactive technology from the aspect of heat flux monitoring or power supplies.

## Materials and Methods

### Materials

Poly(vinylidene fluoride-co-hexafluoropropylene) (PVDF-HFP, Mn = 130,000), lithium tetrafluoroborate (LiBF$_4$, 98 %) and lithium bis(trifluoromethanesulfonyl)imide (LiTFSI) were purchased from Sigma-Aldrich. 1-Ethyl-3-methylimidazolium bis(trifluoromethylsulfonyl)imide (EMIMTFSI), 1-ethyl-3-methylimidazolium tetrafluoroborate (EMIMBF$_4$), 1-ethyl-3-methylimidazolium chloride (EMIMCl), and 1-ethyl-3-methylimidazolium bromide (EMIMBr) were from Aladdin Industrial Corporation. Multi-walled carbon nanotubes are provided by XFNANO. All chemicals were used as received.

**Preparation of the polymer gels**

The polymer gels were synthesized with simple solution process. Firstly, the salts were dissolved in ILs with desired molar concentration in a glove box. Meanwhile, PVDF-HFP was dissolved in acetone, with concentration of 0.1 g mL$^{-1}$. The polymer solution was stirred at 50 °C until PVDF-HFP was completely dissolved in the acetone and the solution became transparent and homogeneous. Then the pre-mixed liquid electrolyte was added to the PVDF-HFP solution. It was stirred for half an hour before use. The solution was then drop casted onto a glass slide and dried in an oven at 60 °C for 10 minutes to form a freestanding polymer gel film.

**Thermopower measurement**

The thermopower measurement was conducted on a homemade setup in the in-plane direction (Fig. S1). Two Peltier devices were used to create hot and cold terminals. Two T-type thermocouples were placed on the copper electrodes near the polymer gel. The thermocouple tips were pasted with thermal grease to ensure accurate measurements of the temperature difference. The Keithley 2182A voltage meter and National Instruments 9213 thermocouples data logger were connected to a computer to record the thermovoltage and temperature every 2 seconds. The thermovoltage usually reached a plateau in 2 minutes. The measurements were conducted at room temperature (~25 °C) and ~60% RH and the humidity influence tests were performed in Terchy humidity chamber (±2.5% RH). We also deposited 50 nm Au protection layer on the copper electrodes and the measured thermopowers were just the same (Fig. S2).

**Material characterizations**

Ionic conductivity for the polymer gels was measured with a stainless-steel blocking electrodes mould, using Metrohm Autolab PGSTAT302N electrochemical workstation for impedance analysis over the frequency range from 0.1 Hz to 1 MHz by applying 10 mV perturbation. The Raman Spectroscopy measurements of the polymer gels were performed using a Renishaw InVia MicroRaman system with 633 nm laser and a Leica objective with 50x magnification. The spectra were fitted by Peakfit v4.12 software. NMR spectra were measured with a Bruker AVII 400 NMR spectrometer at the frequency of 400 MHz using d6-acetone as the solvent. The FTIR spectra were measured using Bruker Vertex 70 Hyperion 1000 spectrometer equipped with the ATR (Attenuated Total Reflection) accessory. The thermal conductivity was measured with hot-disk methods.

**Fabrication of flexible thermoelectric device**

Conductive copper tapes were patterned on a thin polyethylene substrate as the bottom electrodes. A 1-mm-thick 3M VHB tape was punched with 4 mm × 4 mm square pore arrays and then adhered to the bottom electrode layer. Next, copper tapes as the top electrodes were attached to the top of the VHB tape. The p- and n-type ternary polymer solutions with 0.5 M lithium salts or chlorides were injected into the pores using a needle and then dried at 60 °C in an oven. Thereafter, another thin polyethylene film was covered on top. For the thermopower measurement, T-type thermocouples were attached beneath the inner sides of the polyethylene films. During the test, the device was sandwiched between two Peltier chips to generate temperature difference.

**References**

1. C. Forman, I. K. Muritala, R. Pardemann, B. Meyer, Estimating the global waste heat potential. *Renewable and Sustainable Energy Reviews* **57**, 1568–1579 (2016).

2. Z. Sun, J. Lai, S. Wang, T. Wang, Thermodynamic optimization and comparative study of different ORC configurations utilizing the exergies of LNG and low grade heat of different temperatures. *Energy* **147**, 688–700 (2018).

3. G. J. Snyder, E. S. Toberer, Complex thermoelectric materials. *Nature Materials* **7**, 106–114 (2008).

4. A. Boyer, E. Cissé, Properties of thin film thermoelectric materials: application to sensors using the Seebeck effect. *Mater. Sci. Eng. B* **13**, 103–111 (1992).

5. H. Ohta, S. Kim, Y. Mune, T. Mizoguchi, K. Nomura, S. Ohta, T. Nomura, Y. Nakanishi, Y. Ikuhara, M. Hirano, H. Hosono, K. Koumoto, Giant thermoelectric Seebeck coefficient of a two-dimensional electron gas in $SrTiO_3$. *Nat. Mater.* **6**, 129–134 (2007).

6. M. Bonetti, S. Nakamae, M. Roger, P. Guenoun, Huge Seebeck coefficients in nonaqueous electrolytes. *J. Chem. Phys.* **134**, 114513 (2011).

7. D. Zhao, H. Wang, Z. U. Khan, J. C. Chen, R. Gabrielsson, M. P. Jonsson, M. Berggren, X. Crispin, Ionic thermoelectric supercapacitors. *Energy Environ. Sci.* **9**, 1450–1457 (2016).

8. S. L. Kim, H. T. Lin, C. Yu, Thermally chargeable solid-state supercapacitor. *Adv. Energy Mater.* **6**, 1600546 (2016).

9. T. Li, X. Zhang, S. D. Lacey, R. Mi, X. Zhao, F. Jiang, J. Song, Z. Liu, G. Chen, J. Dai, Y. Yao, S. Das, R. Yang, R. M. Briber, L. Hu, Cellulose ionic conductors with high differential thermal voltage for low-grade heat harvesting. *Nat. Mater.* **18**, 608–613 (2019).

10. C. G. Han, X. Qian, Q. Li, B. Deng, Y. Zhu, Z. Han, W. Zhang, W. Wang, S. P. Feng, G. Chen, W. Liu, Giant thermopower of ionic gelatin near room temperature. *Science* **368**, 1091–1098 (2020).

11. H. Cheng, X. He, Z. Fan, J. Ouyang, Flexible quasi-solid state ionogels with remarkable Seebeck coefficient and high thermoelectric properties. *Adv. Energy Mater.* **9**, 1901085 (2019).

12. D. Zhao, A. Martinelli, A. Willfahrt, T. Fischer, D. Bernin, Z. U. Khan, M. Shahi, J. Brill, M. P. Jonsson, S. Fabiano, X. Crispin, Polymer gels with tunable ionic Seebeck coefficient for ultra-sensitive printed thermopiles. *Nat. Commun.* **10**, 1093 (2019).

13. A. Würger, Thermal non-equilibrium transport in colloids. *Reports Prog. Phys.* **73**, 126601 (2010).

14. E. D. Eastman, Theory of the soret effect. *J. Am. Chem. Soc.* **50**, 283–291 (1928).

15. H. Wang, D. Zhao, Z. U. Khan, S. Puzinas, M. P. Jonsson, M. Berggren, X. Crispin, Ionic thermoelectric figure of merit for charging of supercapacitors. *Adv. Electron. Mater.* **3**, 1700013 (2017).

16. J. N. Agar, C. Y. Mou, J. L. Lin, Single-ion heat of transport in electrolyte solutions. A hydrodynamic theory. *J. Phys. Chem.* **93**, 2079–2082 (1989).

17. M. He, F. Qiu, Z. Lin, Towards high-performance polymer-based thermoelectric materials. *Energy Environ. Sci.* **6**, 1352–1361 (2013).

18. S. S. Zhang, K. Xu, T. R. Jow, Low-temperature performance of Li-ion cells with a $LiBF_4$-based electrolyte. *J. Solid State Electrochem.* **7**, 147–151 (2003).

19. J.-C. Lassègues, J. Grondin, D. Talaga, Lithium solvation in bis(trifluoromethanesulfonyl)imide-based ionic liquids. *Phys. Chem. Chem. Phys.* **8**, 5629–5632 (2006).

20. K. Elamin, M. Shojaatalhosseini, O. Danyliv, A. Martinelli, J. Swenson, Conduction mechanism in polymeric membranes based on PEO or PVdF-HFP and containing a piperidinium ionic liquid. *Electrochim. Acta* **299**, 979–986 (2019).

21. K. R. Seddon, A. Stark, M. J. Torres, Influence of chloride, water, and organic solvents on the physical properties of ionic liquids. *Pure Appl. Chem.* **72**, 2275–2287 (2000).

22. Y. Yamada, J. Wang, S. Ko, E. Watanabe, A. Yamada, Advances and issues in developing salt-concentrated battery electrolytes. *Nature Energy* **4**, 269–280 (2019).

23. J. B. Haskins, C. W. Bauschlicher, J. W. Lawson, Ab initio simulations and electronic structure of lithium-doped ionic liquids: structure, transport, and electrochemical stability. *J. Phys. Chem. B* **119**, 14705–14719 (2015).

24. S. A. Katsyuba, P. J. Dyson, E. E. Vandyukova, A. V. Chernova, A. Vidiš, Molecular structure, vibrational spectra, and hydrogen bonding of the ionic liquid 1-ethyl-3-methyl-1H-imidazolium tetrafluoroborate. *Helv. Chim. Acta* **87**, 2556–2565 (2004).

25. J. Tong, X. Xiao, X. Liang, N. v. Solms, F. Huo, H. He, S. Zhang, Insights into the solvation and dynamic behaviors of a lithium salt in organic- and ionic liquid-based electrolytes. *Phys. Chem. Chem. Phys.* **21**, 19216-19225 (2019).

26. C. D'Agostino, M. D. Mantle, C. L. Mullan, C. Hardacre, L. F. Gladden, Diffusion, ion pairing and aggregation in 1-ethyl-3-methylimidazolium-based ionic liquids studied by $^1$H and $^{19}$F PFG NMR: effect of temperature, anion and glucose dissolution. *ChemPhysChem* **19**, 1081-1088 (2018).

27. S. Cha, D. Kim, Anion exchange in ionic liquid mixtures. *Phys. Chem. Chem. Phys.* **17**, 29786–29792 (2015).

28. J. Kiefer, J. Fries, A. Leipertz, Experimental vibrational study of imidazolium-based ionic Liquids: Raman and infrared spectra of 1-ethyl-3methylimidazolium bis(trifluoromethylsulfonyl) imide and 1-ethyl-3-methylimidazolium ethylsulfate. *Appl. Spectrosc.* **61**, 1306–1311 (2007).

29. K. Dong, S. Zhang, D. Wang, X. Yao, Hydrogen bonds in imidazolium ionic liquids. *J. Phys. Chem. A* **110**, 9775–9782 (2006).

30. Y. Kurazumi, T. Tsuchikawa, J. Ishii, K. Fukagawa, Y. Yamato, N. Matsubara, Radiative and convective heat transfer coefficients of the human body in natural convection. *Build. Environ.* **43**, 2142-2153 (2008).

31. Hukseflux, "User manual FHF02SC", https://www.hukseflux.com/uploads/product-documents/FHF02SC_manual_v1803.pdf.

32. A. J. Arduengo, D. A. Dixon, R. L. Harlow, H. V. R. Dias, W. T. Booster, T. F. Koetzle, Electron distribution in a stable carbene. *J. Am. Chem. Soc.* **116**, 6812-6822 (1994).

**Acknowledgments**

The authors are thankful for the financial support from the Hong Kong General Research Fund (Grant Nos. 16206020 and 16214217).

**Author contributions:**

Conceptualization: SL, YY, BH

Methodology: SL, YY

Investigation: SL, YY, HH, JZ, GL,THT

Supervision: BH

Writing—original draft: SL, YY

Writing—review & editing: SL, YY, BH

**Competing interests:** Authors declare that they have no competing interests.

**Data and materials availability:** All data are available in the main text or the supplementary materials.

**Figures**

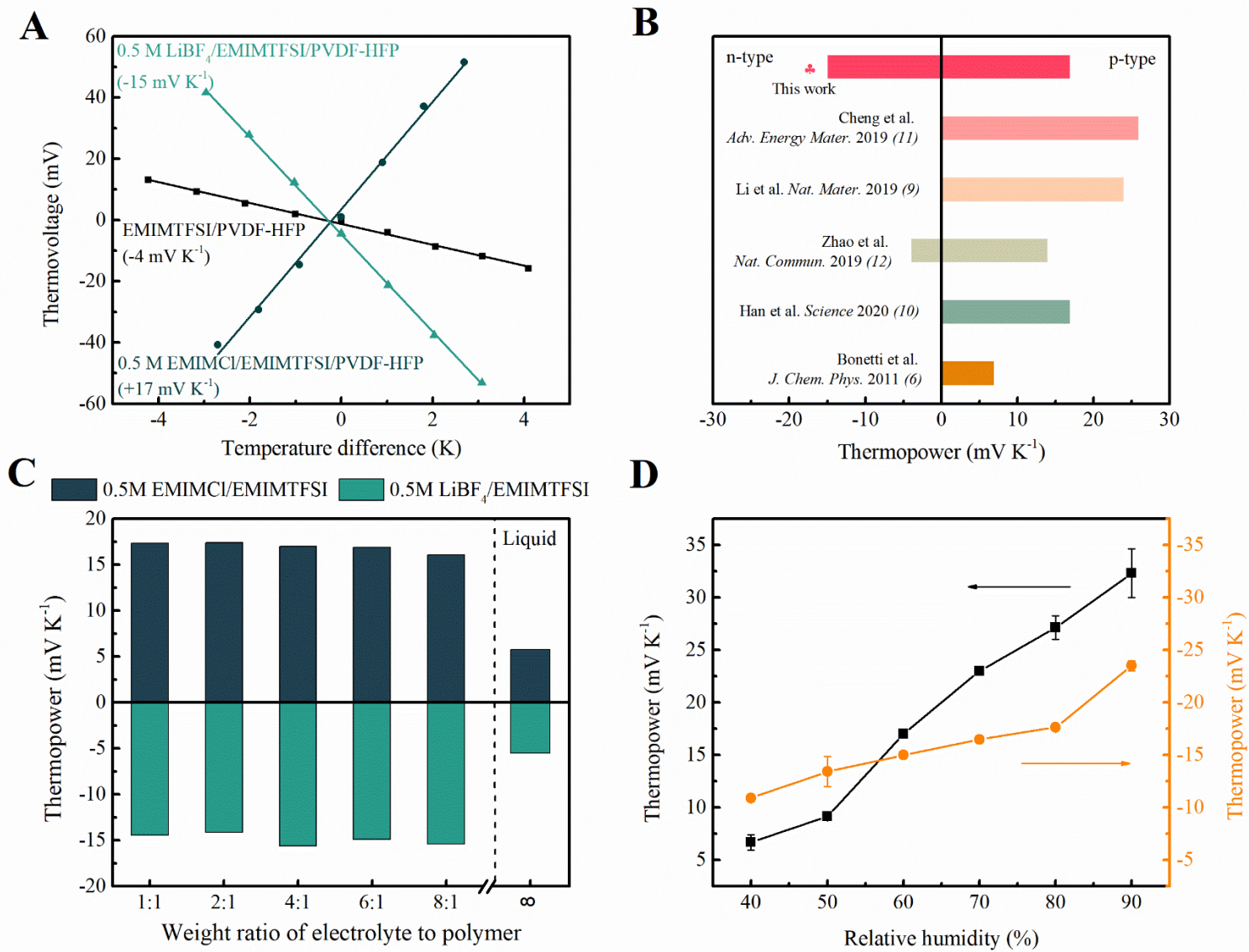

Fig. 1. **Thermopower of the** polymer gels. (A) Thermovoltage of pure EMIMTFSI/PVDF-HFP, 0.5 M EMIMCl/EMIMTFSI/PVDF-HFP and 0.5 M LiBF4/EMIMTFSI/PVDF-HFP ionogels as a function of temperature difference at around 60% RH. The weight ratio of electrolyte to polymer is 4:1. (B) Thermopowers of ionic thermoelectric materials in the literature. (C) Thermopowers of the p- and n-type ionogels for different weight ratios of liquid electrolyte to polymer matrix PVDF-HFP at around 60% RH. The salt concentration is maintained

at 0.5 M. (D) Thermopower of 0.5 M EMIMCl/EMIMTFSI/PVDF-HFP (black squares) and 0.5 M LiBF4/EMIMTFSI/PVDF-HFP (orange circles) ionogels with the dependence of relative humidity.

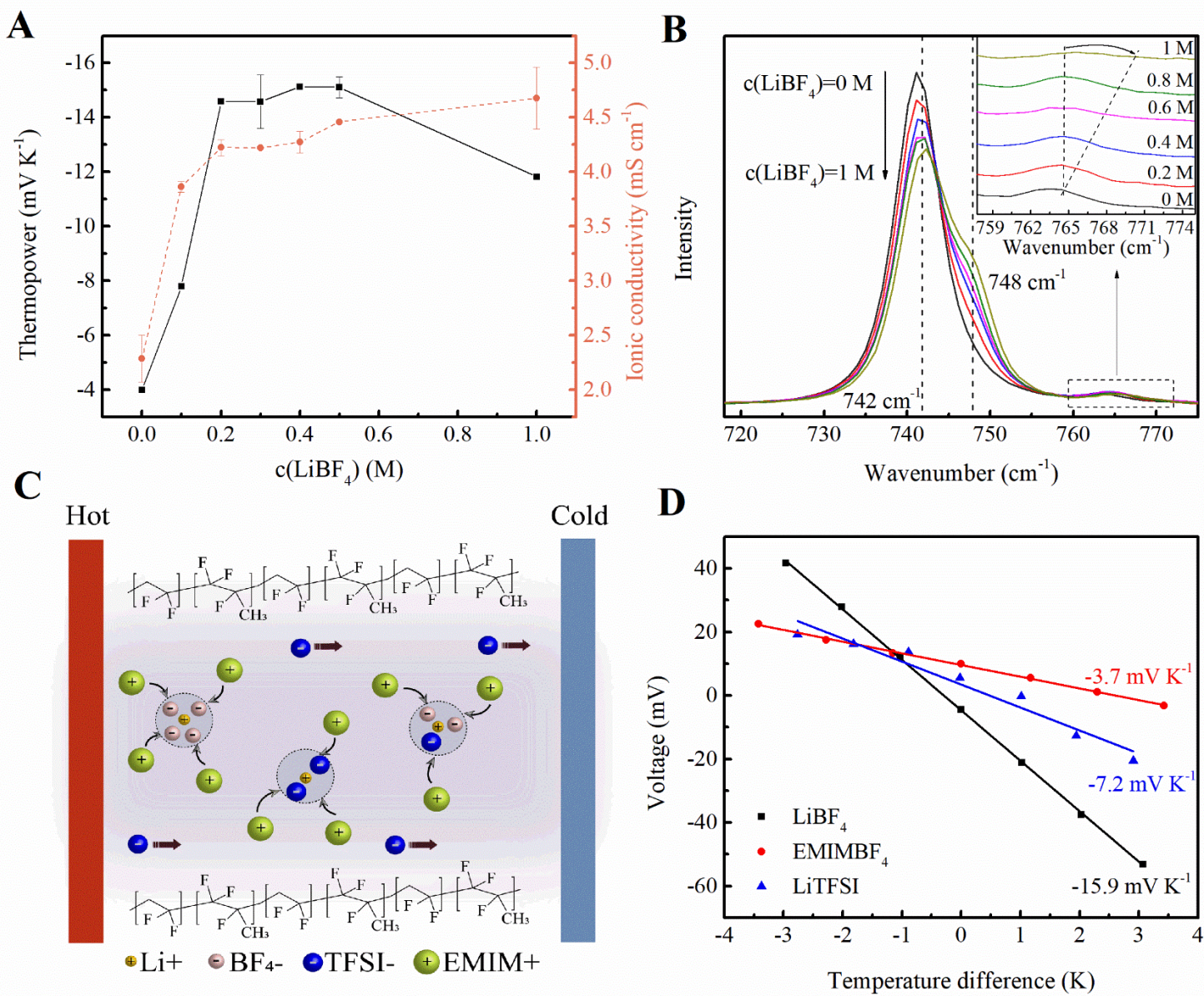

**Fig. 2. Influence of LiBF₄ concentration on the thermoelectric properties of the ionogels.** (**A**) Thermopower and ionic conductivity of the LiBF$_4$/EMIMTFSI/PVDF-HFP ionogel with varying LiBF4 concentration at 60% RH. (B) Raman spectra of LiBF4/EMIMTFSI/PVDF-HFP ionogel for c(LiBF$_4$) = 0, 0.2, 0.4, 0.6, 0.8 and 1 M. All the spectra have been normalized with area between 717 and 775 cm$^{-1}$. The enlarged region between 757 and 775 cm$^{-1}$ is given in the insert for a better vision. (**C**) Schematic to illustrate the effect of lithium clusters on the thermodiffusion of ions in the polymer channel. (**D**) Thermopower of the EMIMTFSI/PVDF-HFP ionogels with 0.5 M LiBF$_4$, EMIMBF$_4$ and LiTFSI, respectively.

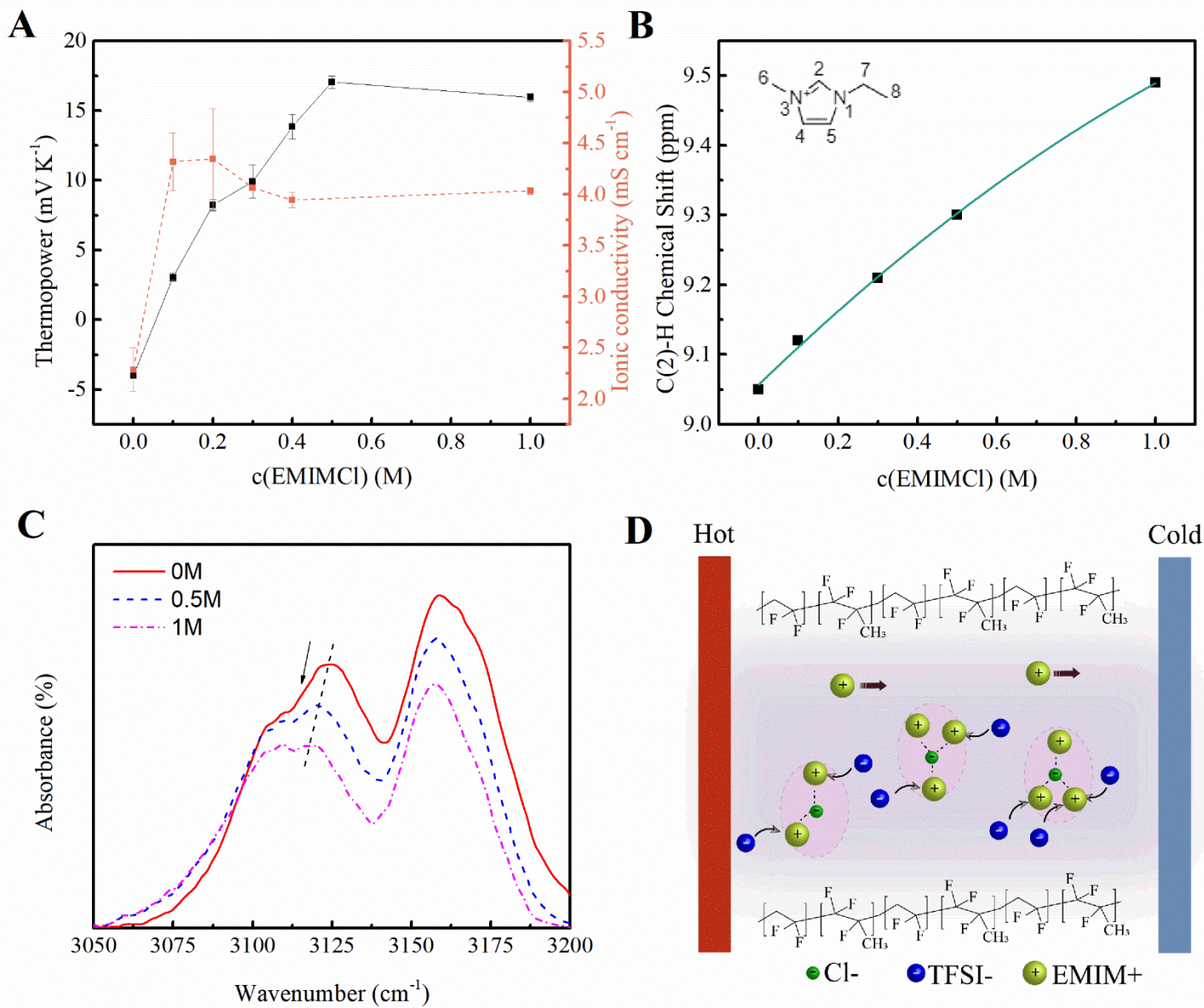

**Fig. 3. Influence of EMIMCl concentration on the thermoelectric properties of the ionogels.** (**A**) Thermopower and ionic conductivity of the EMIMCl /EMIMTFSI/PVDF-HFP ionogel with varying EMIMCl concentration at 60% RH. (**B**) Chemical shift of C(2)-H in the EMIMCl/EMIMTFSI mixture. (**C**) FTIR spectra of the EMIMCl/EMIMTFSI/PVDF-HFP ionogel, c(EMIMCl) = 0, 0.5 and 1 M. (**D**) Schematic to illustrate the effect of ion aggregates on the thermodiffusion of ions in the polymer channel.

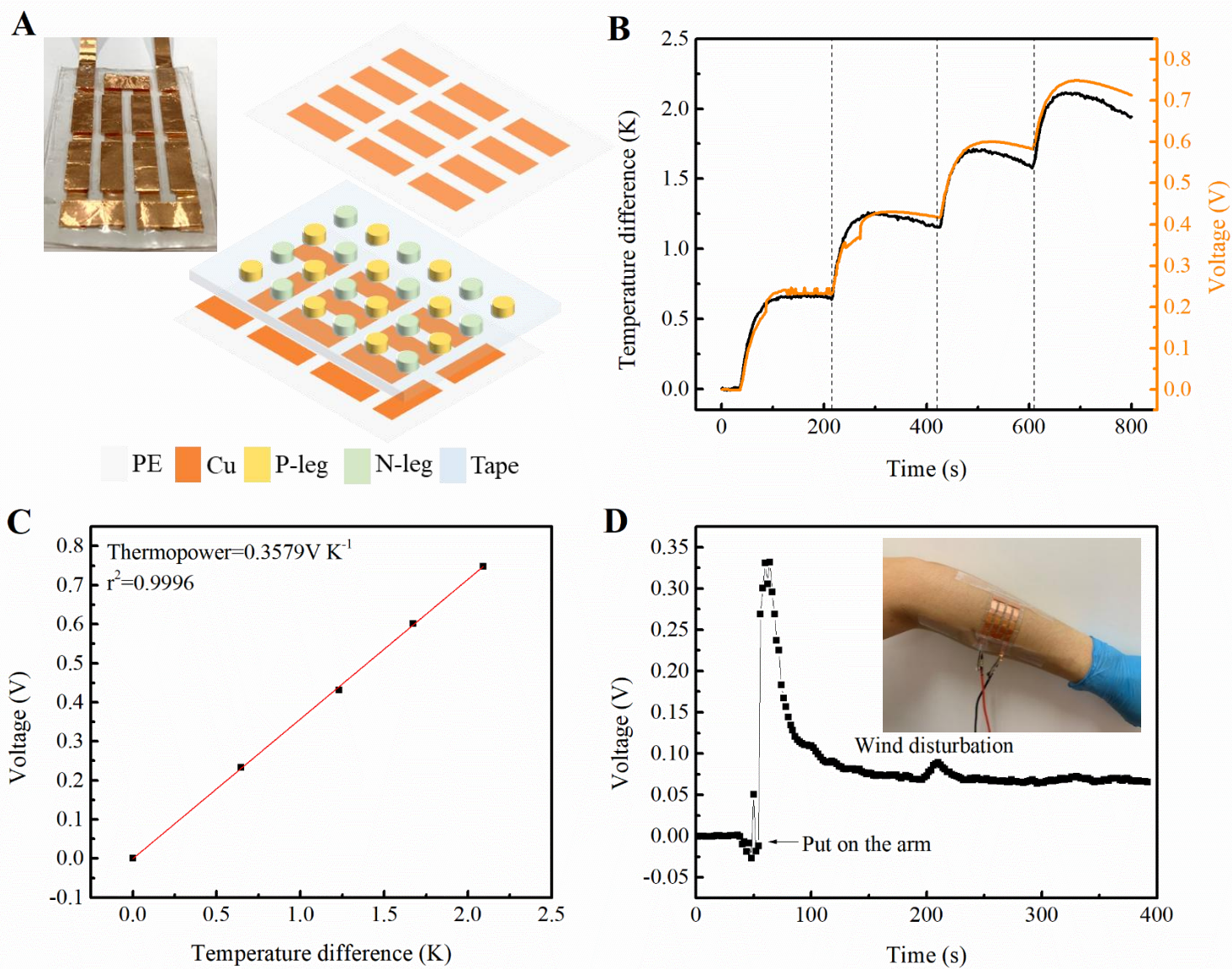

**Fig. 4. Flexible thermoelectric device.** (**A**) A schematic of the design with 12 pairs of p- and n-type ionogels together with the picture of the device. (**B**) Voltage changes with temperature difference. (**C**) Linear fitting of voltage to temperature difference. (**D**) Voltage generated from the device when put on a human arm at 25 °C.

**Supplementary Materials**

Figs. S1 to S8

Tables S1